\documentclass[12pt]{article}
\usepackage{amsmath}
\usepackage{amssymb}
\usepackage[dvips]{graphicx}
\usepackage[dvips]{color}

\begin{document}
\title{Subthreshold oscillations in a map-based neuron model}

\author{
ANDREY L. SHILNIKOV\\ {\em Department of Mathematics and
Statistics,}\\ {\em Georgia State University, Atlanta, GA
30303-3083, USA}\\
\and\\
NIKOLAI F. RULKOV \\ {\em Institute for Nonlinear Science, UCSD}\\
{\em San Diego, CA  92093-0402, USA} }

\date{\today}
\maketitle
\begin{abstract}
Self-sustained subthreshold oscillations in a discrete-time model
of neuronal behavior are considered. We discuss bifurcation
scenarios explaining the birth of these oscillations and their
transformation into tonic spikes. Specific features of these
transitions caused by the discrete-time dynamics of the model and
the influence of external noise are discussed.
\end{abstract}

\vskip1pc
\newcounter{eq}

\section{Introduction}

Studies of dynamical behavior of biological networks require
numerical simulations of arrays containing a very large number of
neurons. Despite the variety of physiological processes involved in
the formation of neuron activity, the thorough studies of the
large-scale networks need simple phenomenological models that can
replicate the dynamics of individual neurons. Various suggestions
for the design of low-dimensional maps for modeling the neurons'
behavior have been proposed, see for
example~\cite{Chiavlo93,Chailvo95,Cazelles01,Rulkov01,Andreev02,IH03}
and references therein. Most of them were focused on the
replication of either fast spikes or relatively slow bursts while
the mechanisms for generation of specific footprints of spikes were
neglected.

A simple discrete-time model replicating the spiking-bursting
neural activity has been suggested recently in~\cite{Rulkov2002}.
This model is a 2D map that mimics rather realistically various
types of transitions that occur in biological neurons. These
transitions include routes between silence and tonic spiking as
well as a triplet: silence $\leftrightarrow$  bursts of spikes
$\leftrightarrow$  tonic spiking. Such simple phenomenological
models bear a high potential for further developments of
computationally efficient methods for studies of functional
behavior in large-scale neurobiological networks~\cite{rbsNWP03}.

The bifurcation analysis of the map model carried out in
\cite{srIJBC2003} has shown that the transition from silence (a
stable fixed point) to generation of action potentials is
characterized by a sub-critical Andronov-Hopf bifurcation when an
unstable invariant closed curve collapses into the stable fixed
point. Therefore, the original map-model \cite{Rulkov2002} provides
only an abrupt transition from silence to spiking as a control
parameter (e.g. the depolarization current) passes the excitability
threshold. This scenario is quite typical for  most types of
biological neurons. However, experimental studies suggest that some
neurons may come out of the silence softly through the regime of
small oscillations below the threshold of the spike
excitation~\cite{Llinas88}. These subthreshold oscillations of
almost sinusoidal form facilitate the generation of spike
oscillations when the membrane gets depolarized or
hyperpolarized~\cite{Llinas81,Llinas86}. These small oscillations
can play an important role in shaping specific forms of rhythmic
activity that are vulnerable to the noise in the network
dynamics~\cite{Makarov01,Velarde01}.

In this paper we modify the map model so that it can generate
stable subthreshold oscillations. We start the discussion of the
model dynamics with the analysis of local bifurcations of a fixed
point of the map. We show that the loss of stability of the fixed
point is accompanied by the birth of the stable invariant circle
which initiate a family of canards in the map. Further evolution of
the circle leads to a breakdown of the invariant circle that gives
rise to chaos. We also elaborate on the mechanism of the onset of
irregular spiking which is due to heteroclinic-like crossings
between the stable and unstable invariant sets, which are the
images of the slow motion ``surfaces'' in the unperturbed map. The
role played by small subthreshold oscillations in the
responsiveness of the map to external noise is considered.

\section{Map-based model with stable subthreshold oscillations}\label{Section2}

\begin{figure}[ht]
\centering\includegraphics[scale=0.4]{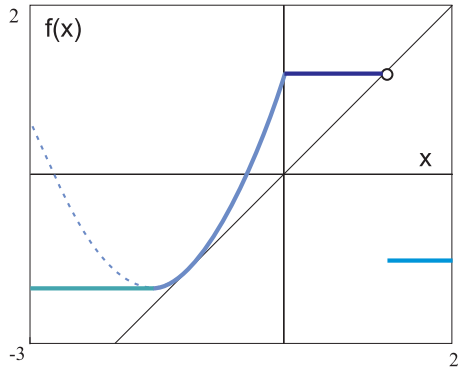} \caption{Geometry
of the fast subsystem, map~(\ref{mapx}), for the parameter values
at the tangent bifurcation: $\alpha=1$ and $y=0$. The function is
discontinuous at the point $x=y+1+ \beta$ which belongs to the
rightmost interval, see (\ref{func}). } \label{fig-rhs}
\end{figure}

The map-based model of spiking-bursting neuron oscillations,
following \cite{Rulkov2002}, can be written in the form of the
two-dimensional map
 \setcounter{eq}{\value{equation}}
\addtocounter{eq}{+1} \setcounter{equation}{0}
\renewcommand{\theequation}{\theeq \alph{equation}}
\begin{eqnarray}
\bar x &=& f_{\alpha}(x,y+\beta), \label{mapx}\\
\bar y &=& y- \mu (x+1-\sigma), \label{mapy}
\end{eqnarray}\label{map1}\newcommand{\refmap}{1}
where the $x$-variable replicates the dynamics of the membrane
potential, the parameters $\alpha$, $\sigma$ and $0<\mu\ll 1$
control individual dynamics of the system. Some input parameters
$\beta$ and $\sigma$ are employed to provide coupling with other
such models afterwards; both stand for injected currents. The
principal distinction of the original map analyzed
in~\cite{Rulkov2002,srIJBC2003} and the one in question is
camouflaged in the function $f_{\alpha}(x,y+\beta)$, which is
given by \setcounter{equation}{\value{eq}}
\renewcommand{\theequation}{\arabic{equation}}
\begin{eqnarray}
f_{\alpha}(x,y+\beta)=\left \{
\begin{array}{ll}
-\alpha^2/4-\alpha+y+\beta,  &  x < -1-\alpha/2, \\
 \alpha x+(x+1)^{2}+y+\beta, & -1-\alpha/2 \leq x \leq 0, \\
 y+1+\beta,                  &  0<x<y+1+ \beta, \\
 -1,                         & x \geq y+1+\beta.
 \end{array} \right .
\end{eqnarray}\label{func}
The graph of this function is pictured in Fig.~\ref{fig-rhs}. The
shape of nonlinear function is meant to achieve a replication of
sharp tonic spikes in the dynamics of the $x$-variable in the map.
The slow $y$-variable can turn the spike generator on and off. The
main difference between the function (\ref{func}) and that in the
map proposed in~\cite{Rulkov2002} is its shape on the left hand
side of the discontinuity point, i.e. at $x < y+1+\beta$. Now the
function contains an interval of parabola instead of a hyperbola
used in \cite{Rulkov2002,srIJBC2003}. This modification is crucial
for stability of subthreshold oscillations. The parabola reaches
its minimum at $x=-1-\alpha/2$. At this point the graph of the
function is continued leftward by a horizontal line, see
Fig.~\ref{fig-rhs}.

\subsection{Fast and slow motions of the map}\label{fast_slow}

\begin{figure}[ht]
\centering\includegraphics[scale=0.4]{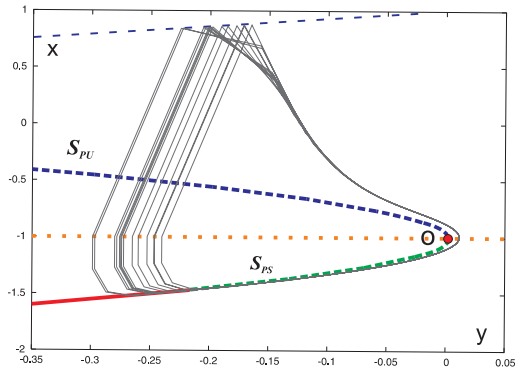} \caption{
$(y,x)$-bifurcation diagram of the fast subsystem,
map~(\ref{mapx}), for $\alpha=1$ and $\mu=0$. The branches $S_{pu}$
and $S_{ps}$ are traced out by the stable and unstable fixed points
of the fast map as the parameter $y$ varies. When $0<\mu \ll 1$,
this $(y,x)$ plane becomes the phase portrait of the 2D map. The
shown trajectory of the 2D map is computed at $\mu=0.04$,
$\alpha=0.99$ and $\sigma=0$. The dashed curves are the zero level
ones for function~(\ref{func}).} \label{yx1}
\end{figure}

For map~(\refmap), when $\mu=0$,  the slow subsystem~(\ref{mapy})
is decoupled from the fast subsystem~(\ref{mapx}), in which $y$ is
regarded as a perturbation parameter. One may see from
Fig.~\ref{fig-rhs} that depending on $y$, the fast subsystem may
have two fixed points, one stable and one unstable, or no fixed
points. The transition between these states occurs via a
saddle-node bifurcation. When $y$ is varied, the fixed points trace
out a parabola in the $(y,x)$-plane as shown in Fig.~\ref{yx1}. The
traces of stable and unstable fixed points form on the $(x,y)$
stable, $S_{ps}$, and unstable, $S_{pu}$, branches, respectively.

The point of intersection of these branches with the nullcline of
slow subsystem (\ref{mapy}), which is given by $x_{fp}=\sigma-1$,
is a fixed point of the two-dimensional map (\refmap). It is easy
to see that this fixed point $O$ is stable if it is located on
$S_{ps}$ and is unstable if it is on $S_{pu}$. The case where the
nullcline crosses the parabola at the fold point requires a more
delicate analysis.

The nonstandard analysis~\cite{arnold94} predicts that when $0<\mu
\ll 1$, the normally hyperbolic branches $S_{ps}$ and $S_{ps}$
persist in the form of the stable and unstable ``slow motion" sets
${\cal S}_{s}$ and ${\cal S}_{u}$ that remain $\mu$-close to the
originals, but $\mu^{3/2}$-close to them near the fold~\cite{F2}.

The idea behind generation of tonic spikes in the map (\refmap) is
illustrated in Fig.~\ref{yx1}. The period of spiking is comprised
of the following phases: the rest phase, where the phase point
slides along the set ${\cal S}_{s}$ at a rate of order $\mu$
towards the fold point. Then, the phase point jumps up indicating
the beginning of a spike. Its upward motion is stopped by a
delimiter (the third segment of function~(\ref{func})) that
reflects the phase point towards the stable slow surface ${\cal
S}_{s}$ through the line segment $x=-1$.

\subsection{Birth of invariant curve}\label{Section4}
Next we carry out the bifurcation analysis of the fixed point $O$.
Our consideration is restricted to the domain
$\alpha^2/2-\alpha+y+\beta \le x<0$, i.e. to the parabolic segment
of function~(\ref{func}). This means that the fast
subsystem~(\ref{mapx}) is chosen to be close to the tangent
bifurcation. The moment of the bifurcation is pictured in
Fig.~\ref{fig-rhs}. From (\ref{mapy}) one finds the $x$-coordinate
of the fixed point $x_{fp}=\sigma-1$. Therefore, the fixed point is
located at the parabolic segment of function when the parameter
values are within the range $-\alpha/2<
\sigma <1$. The $y$-coordinate of the fixed point for this range
of parameters is $y_{fp}=(\sigma-1)(1-a)-\sigma^{2}-\beta$.

Since fixed point $O$ is a single fixed point of map (\refmap), the
further stability analysis is reduced to two possible local
bifurcations: a flip where one of the multipliers of the fixed
point equals $-1$; and the Andronov-Hopf bifurcation, where the
fixed point possesses a pair of multipliers equal $e^{\pm i
\varphi}$ on the unit circle. Simple calculations reveal that the
flip bifurcation takes place outside of the considered parameter
region.

In the case of the Andronov-Hopf (AH) bifurcation, the Jacobian
$$ \bf J= \left [
\begin{array}{cc} \alpha+2\sigma & 1\\
-\mu & 1
\end{array}
 \right ]
$$
of the map equals $1$ at the fixed point, while its trace equals
$2\cos \varphi$. The equation of the corresponding bifurcation
curve $AH$ is given by
\begin{equation}\label{AH}
\alpha_{AH}=-2\sigma+1-\mu.
\end{equation}
On this curve, the fixed point has a pair of complex conjugate
multipliers:
\begin{equation}\label{mult1}
\rho_{1,2}=1-\frac{\mu}{2} \pm \frac{i}{2}\sqrt{\mu(4-\mu)} = \cos
\psi \pm i \sin \psi.
\end{equation}

To determine the stability of fixed point $O$ right at the
bifurcation state we need to evaluate the sign of the first
Lyapunov coefficient $L_{1}$. Note that the value of  $L_{1}$ is
not an invariant as it depends on coordinate transformations. The
critical fixed point $O$ is stable if $L_{1}<0$, and unstable if
$L_{1}>0$.

\begin{figure}[ht]
\centering\includegraphics[scale=0.6]{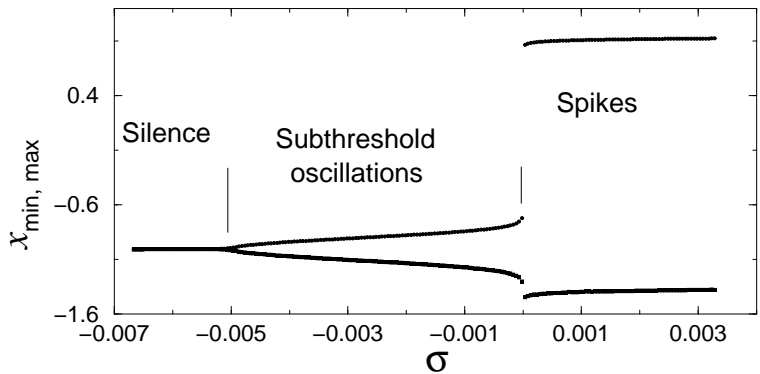}
\caption{Bifurcation diagram illustrating the birth of ``small"
subthreshold oscillations transforming into spikes as parameter
$\sigma$ increases. Top and bottom branches corresponds to the
highest and lowest values of the $x$-variable for a given value of
parameter $\sigma$. The other parameters of the map are set as
follows $\alpha=0.99$, $\beta=0.0$ and $\mu=0.02$.}
\label{bifd}
\end{figure}

Let us introduce new coordinates in which the fixed point is
translated to the origin:
$$
\left ( \begin{array}{c} x \\ y
\end{array} \right )
\mapsto \left ( \begin{array}{cc}
 x +1 - \sigma\\
y + (\sigma-1)(\alpha-1)+\sigma^{2}+\beta
 \end{array}
\right ).
$$
Now the map looks as follows
\begin{equation}\label{mapfp1}
 \bar x = (1-\mu)x+x^{2}+y, \qquad  \bar y= y-\mu x.
\end{equation}
Let us next make another transformation
 $$
\left ( \begin{array}{c} x \\ y \end{array} \right ) \mapsto \left
( \begin{array}{cc}
 0 & 1 \\
 \sin \psi  &  1 -\cos \psi  \end{array}
\right ) \left ( \begin{array}{c} \xi \\ \eta \end{array} \right
),
 $$
with $ \sin \psi$  and $\cos \psi$  defined in (\ref{mult1}). This
 makes the linear part of (\ref{mapfp1}) a rotation
through $\psi$:
 $$
\left ( \begin{array}{c} \bar \xi \\
\bar \eta \end{array} \right )
  =
\left ( \begin{array}{cr}
 \cos \psi  & - \sin \psi  , \\
 \sin \psi  &   \cos \psi,
 \end{array} \right )
\left ( \begin{array}{c} \xi \\
 \eta \end{array} \right ) +
\left ( \begin{array}{c}
\frac{\cos \psi -1}{\sin \psi}  \\
1
 \end{array} \right )\eta^{2}.
 $$
In variable $z=\xi + i\eta$  the map assumes the complex form
\begin{equation}\label{complex}
\bar z = z e^{i \psi}+\frac{c_{20}}{2}z^{2}+c_{11}z z^{*} +
\frac{c_{02}}{2}{z^{*}}^{2}
\end{equation}
with
 \begin{equation}\label{coeffs}
 c_{20}= \frac{1}{2} \left (\tan\frac{\psi}{2}-i \right ),
 \quad c_{11}= \frac{1}{2} \left (i-\tan\frac{\psi}{2} \right ),
 \quad c_{02}= \frac{1}{2} \left (\tan\frac{\psi}{2}-i \right ).
 \end{equation}
The normalizing transformation
 $$
z \mapsto z- \frac{c_{20}}{e^{2i\psi}-e^{i\psi}}z^{2}-
\frac{c_{11}}{1-e^{i\psi}}z z^{*} -
\frac{c_{02}}{e^{-2i\psi}-e^{i\psi}}{z^{*}}^{2}
 $$
eliminates all the quadratic terms in (\ref{complex}) so that the
coefficient $L_{1}$ at the desired cubic term in the resulting
normal form
$$
\bar{z}=e^{i\psi}z(1+(L_1+iS_1)z z^*) + O(\|z\|^3)
$$
becomes the sought Lyapunov value. Its expression reads as
follows:
 \begin{equation}\label{lyapunov1}
L_{1}= -{\mbox {Re}} \frac{(1 -2 e^{i \psi}) e^{-2 i \psi} c_{20}
c_{11}} {2(1 -e^{i \psi})} - \frac{|c_{11}|^2}{2} -
\frac{|c_{02}|^2}{4} .
 \end{equation}
Substituting (\ref{coeffs}) into (\ref{lyapunov1}) yields
 $$
L_1=-\frac{1}{4}\frac{\cos(\psi)}{\cos(\psi)+1}=-\frac{2-\mu}{4(4-\mu)}
<0
$$
for all small $\mu$. Thus, the loss of stability of fixed point
$O$ on the Andronov-Hopf bifurcation curve is accompanied by the
birth of a stable closed invariant curve emerging from $O$. This
mechanism of the birth of subthreshold periodic oscillations is
illustrated in Fig.~\ref{bifd} as the parameter $\sigma$
increases.

\begin{figure}[h]
\centering\includegraphics[scale=0.5]{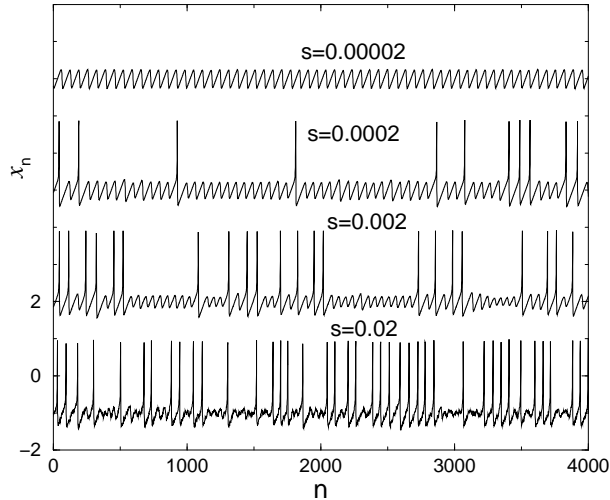} \caption{Four
waveforms of iterates $x_n$ versus the discrete time $n$ show the
influence of an external Gaussian white noise with standard
deviation $s$ on the map oscillations. The control parameters of
the map are $\alpha=0.99$, $\sigma=-0.0001$ and $\mu=0.02$.}
\label{waveforms}
\end{figure}
\begin{figure}[h]
\centering
\includegraphics[scale=0.4]{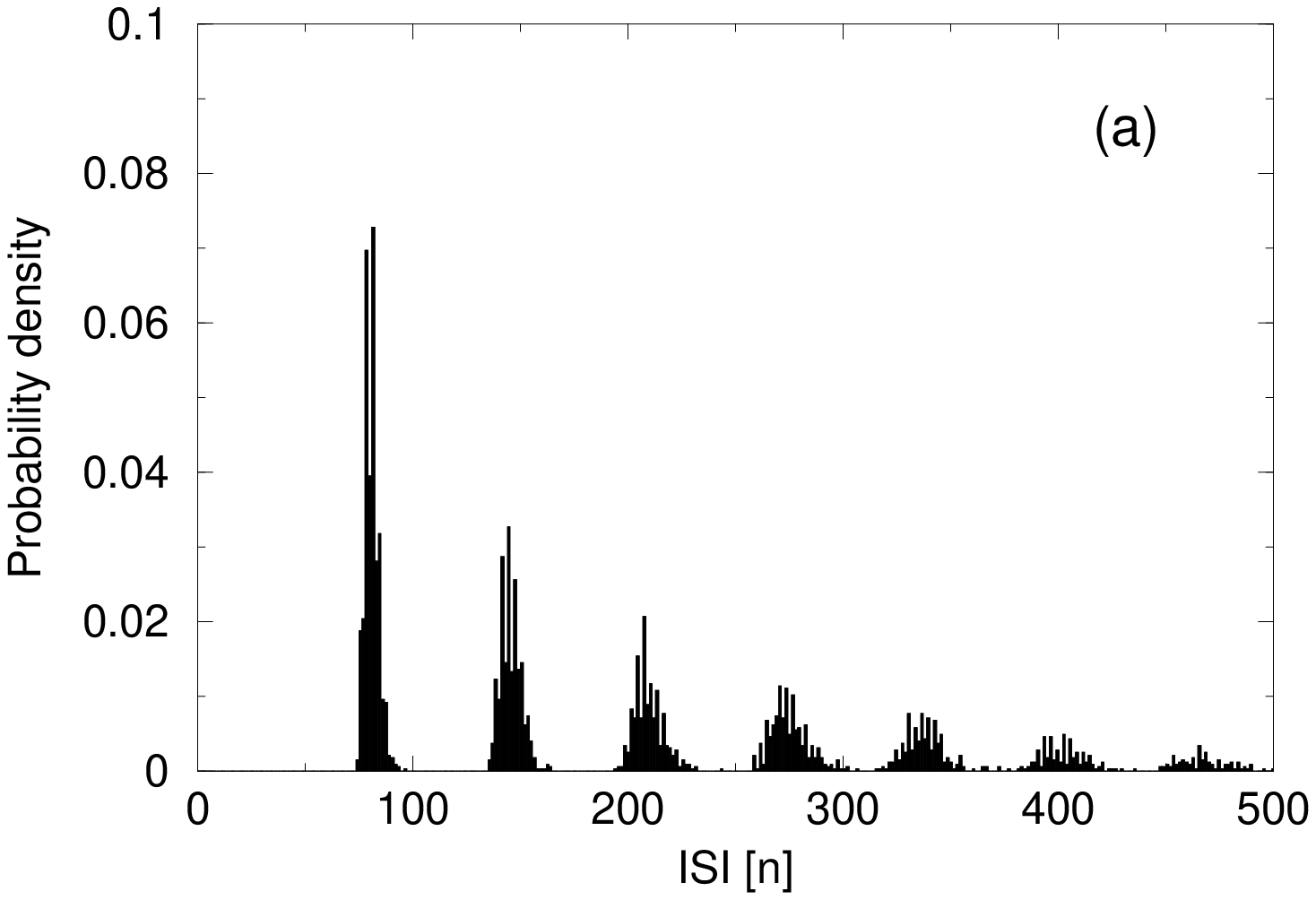}
\includegraphics[scale=0.4]{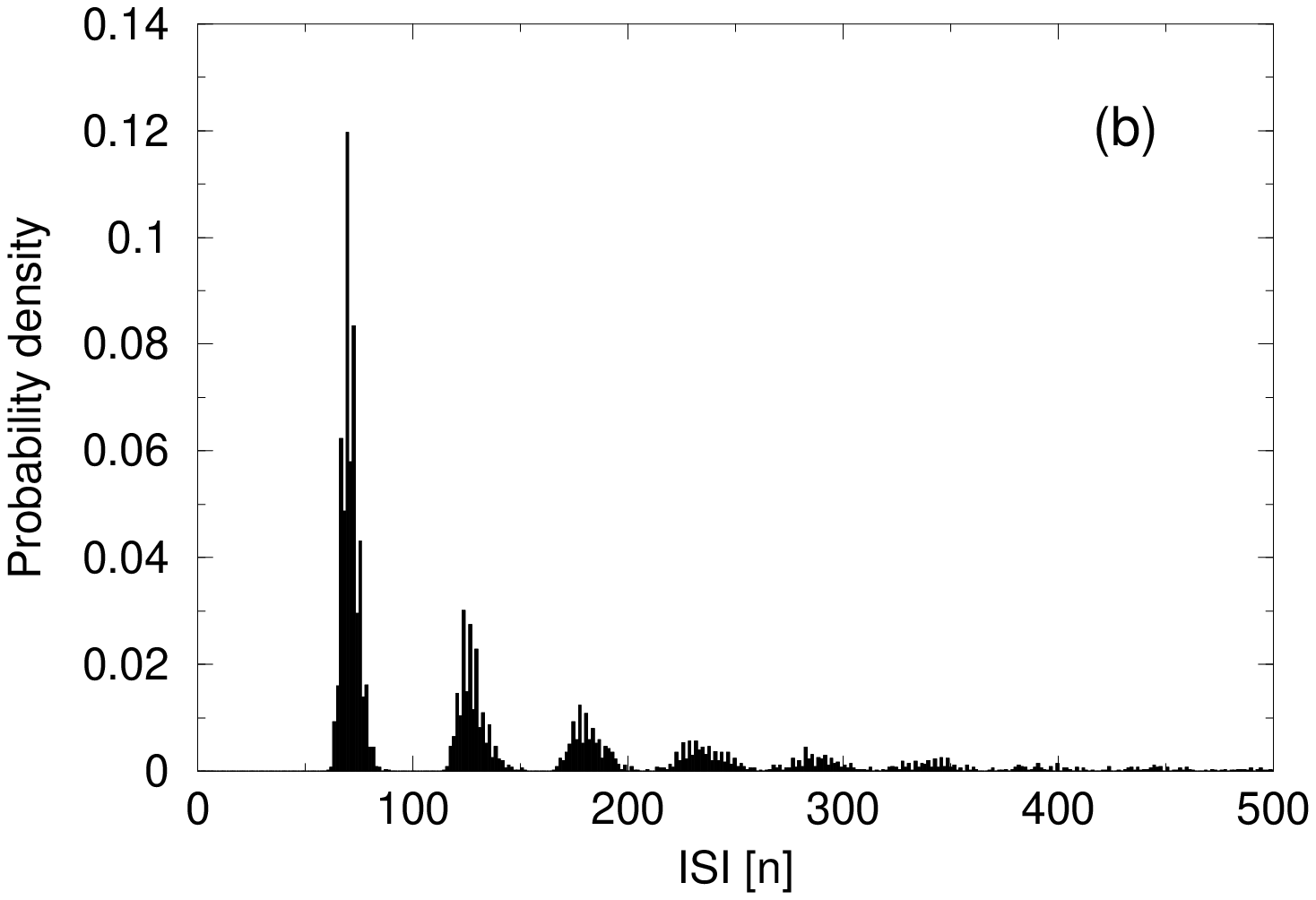}
\includegraphics[scale=0.4]{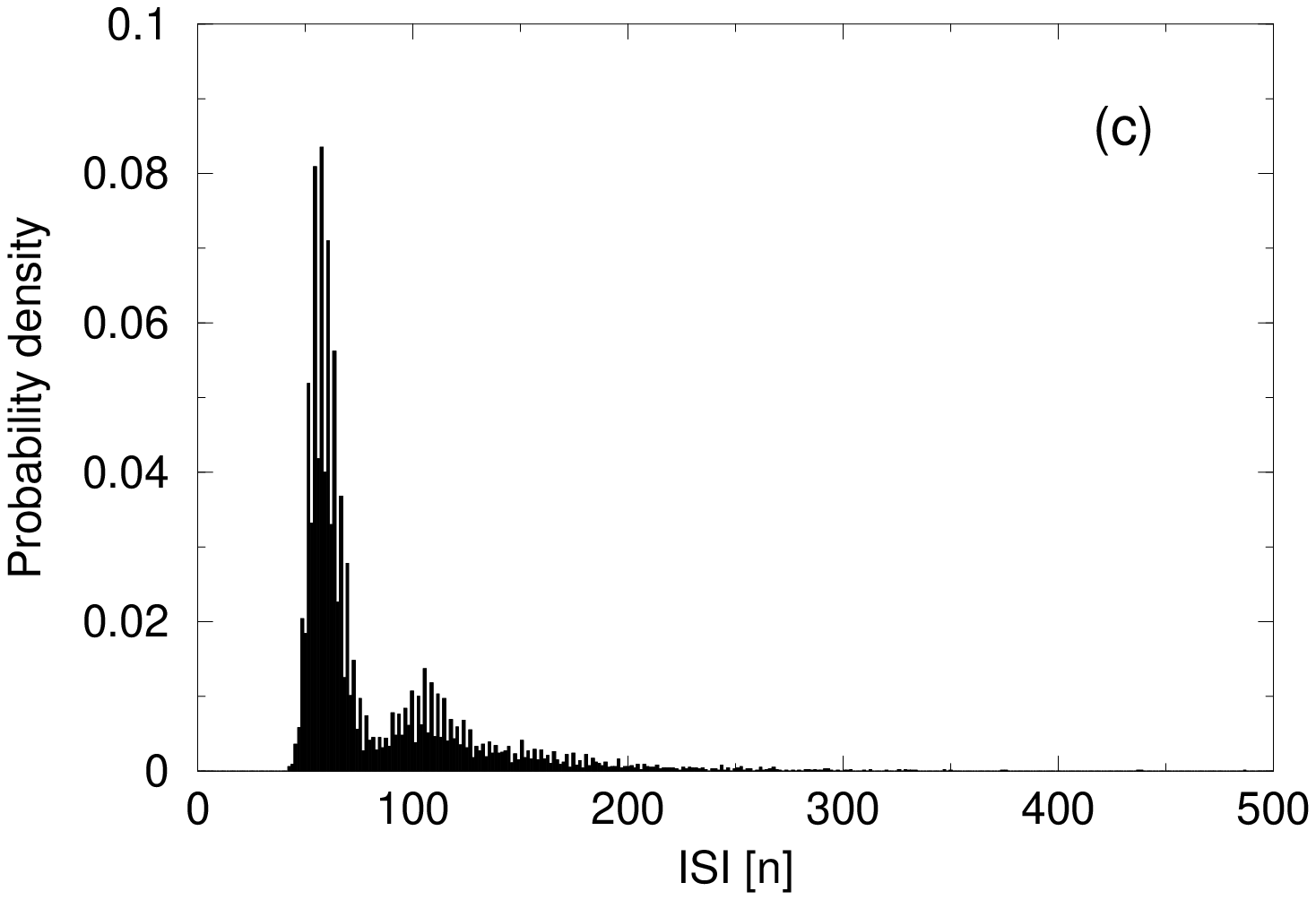}
\caption{Histograms of the interspike interval (ISI) distribution
computed for the waveforms shown in Fig.~\ref{waveforms}. (a) -
$s=0.0002$; (b) - $s=0.002$ and (c) - $s=0.02$.} \label{hist}
\end{figure}

\subsection{Noise and subthreshold oscillations}
It was observed recently that subthreshold periodic activity shapes
stochastic properties of spiking in a neuron influenced by noise,
see for
example~\cite{Braun,Heinz,Makarov01,Velarde01,Chik01,Volkov03}. To
illustrate similar properties in our model let us now consider the
regime of subthreshold oscillations in the map with the presence of
noise. For the sake of briefness we consider only the case when
noise is applied to the fast subsystem. The discussion of the
effects of noise, occurring in the fast and slow subsystems, can be
found elsewhere~\cite{Hilborn1,Hilborn2}. The system~(\refmap) in
presence of noise can be written as follows
\begin{eqnarray}
 x_{n+1} &=& f_{\alpha}(x_n,y_n+\beta)+\zeta_n, \label{mapxn}\\
 y_{n+1} &=& y_n- \mu (x_n+1-\sigma), \label{mapyn}
\end{eqnarray}
where $\zeta_n$ is a delta-correlated Gaussian White Noise (GWN)
with zero mean value and standard deviation value $s$.

Waveforms of map model (\refmap) operating in the regime of
subthreshold oscillations and the influence of noise are shown in
Fig.~\ref{waveforms}. Four traces plotted in the figure correspond
to different values of the standard deviation, $s$, of GWN. The top
trace presents the case where the level of noise is insufficient to
induce an action potential (a spike). When the level of noise
exceeds a critical level the map starts producing occasional
spikes. Such spikes are more likely to occur at the top of
oscillation. This behavior results in the formation of a multi-hump
structure in the probability distribution function of interspike
intervals (ISIs), see histograms shown in Fig.~\ref{hist}(a,b).

A further raise of the noise level increases the probability of the
action potentials. As the result, spikes occur almost every period
of the subthreshold oscillation (see the bottom trace in
Fig.~\ref{waveforms}) and the distribution of probability density
of ISIs transforms into a single hump structure, see
Fig.~\ref{hist}(c). These results are in good agreement with the
results obtained from ODE based neuron
models~\cite{Chik01,Volkov03}.

\section{Tangles of critical curves and chaos}\label{Section6}

\begin{figure}[ht]
 \centering\includegraphics[scale=0.8]{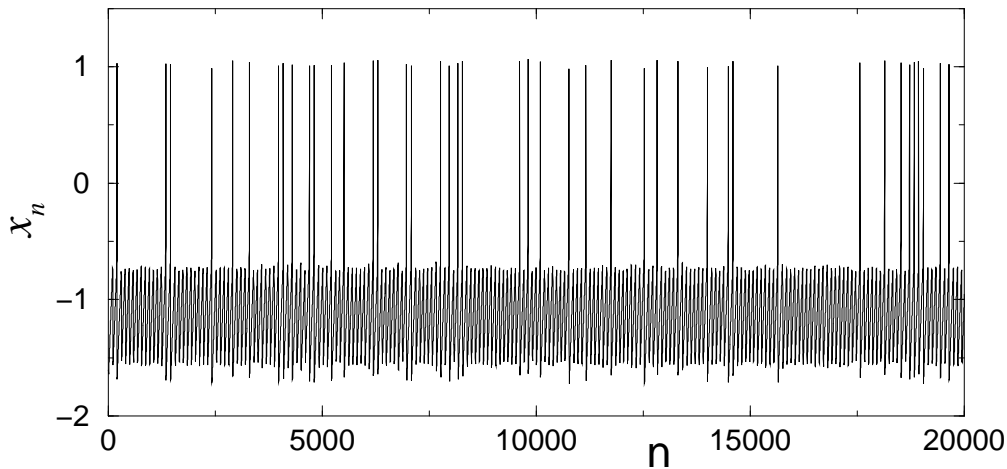}
\caption{ Sporadic spiking in the map~(\refmap) computed with
$a=1.25$, $\sigma=-0.13$ and $\mu=0.02$. The phase portrait of
these oscillations is shown in Fig.~\ref{chaos}(a).}\label{chaos1}
\end{figure}

In the numerical simulations of map~(\refmap) we found that
subthreshold oscillations may be interrupted by irregular spiking
even in the absence of external noise. An example of such a
behavior is presented in Fig.~\ref{chaos1}. This intermediate
dynamics is observed only within a rather thin parameter interval
at the border between the regimes of continuous subthreshold
oscillations and tonic spike generation. To understand the
dynamical mechanisms behind this sporadic spiking we numerically
studied the evolution of an invariant circle as the parameter
values enter this thin region.

It follows from the theory of canards that the parameter domain for
the existence of a stable invariant circle is a narrow strip of the
order of $O(\mu)$ that adjoins the bifurcation curve $AH$.
Furthermore, the size of the circle increases abnormally fast as
the parameter values deviate from the Andronov-Hopf bifurcation
curve and approach the critical values where the invariant circle
breaks down. Such extreme sensitivity to slight parameter
deviations hamper the detailed analysis of bifurcations associated
with the circle as it breaks.

\begin{figure}[ht]
 \centering\includegraphics[scale=0.4]{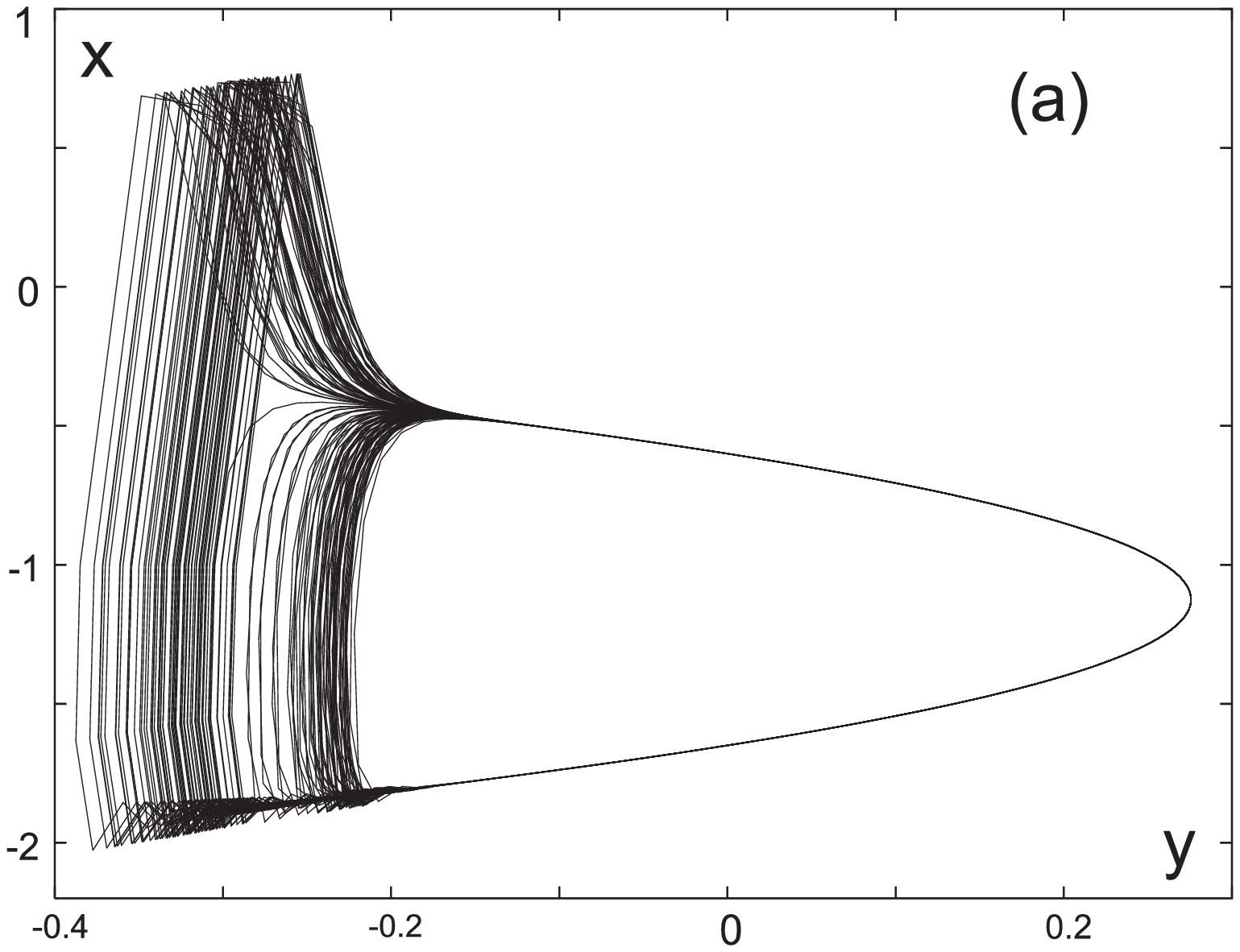}\qquad
 \includegraphics[scale=0.42]{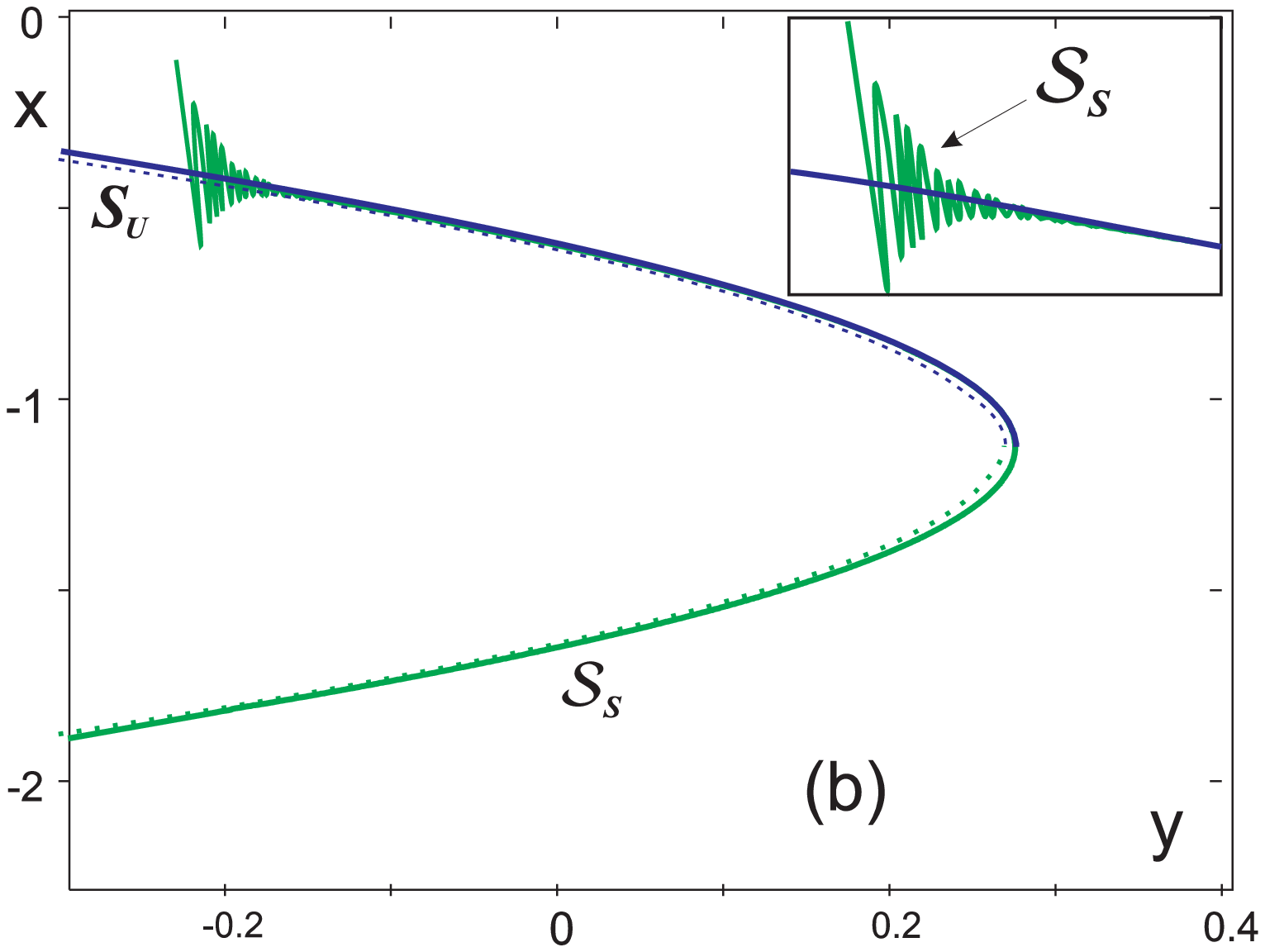}
\caption{(a) Chaos in the noiseless map~(\refmap) computed with
$\alpha=1.25$, $\sigma=-0.13$ and $\mu=0.02$ The corresponding
waveform is shown in fig.~\ref{chaos1}. (b) Forward iterates of a
small interval of the stable critical set ${\cal S}_{s}$ reveal
increasing wiggles occurred around the unstable critical set ${\cal
S}_{u}$. }\label{chaos}
\end{figure}

The breakdown of the stable invariant circle leads to an
interesting situation depicted in Fig.~\ref{chaos}(a). One can
observe the co-existence of two kinds of special solutions
following the unstable slow branch ${\cal S}_{u}$. They are called
canards with a head (a spike) and ones without it \cite{arnold94}.
A canard is characterized by a growing level of exponential
instability with respect to nearby solutions. This instability is a
necessary first component of chaotic behavior observed in a system.
In addition, the presence of two types of canards creates mixing
and uncertainty, which is the second important ingredient for the
onset of chaos. This type of chaos in the neuron model (\refmap)
appears as small subthreshold oscillations alternating with
sporadic spikes, see Fig.~\ref{chaos1}.

To understand the dynamical mechanism behind the splitting of
canards into two types, we conducted a numerical analysis of the
behavior of ${\cal S}_{s}$ and ${\cal S}_{u}$ for various parameter
values selected close to the threshold for the breakdown of the
stable invariant circle. To plot ${\cal S}_{s}$, we iterated
forward a large number of phase points initiated from a relatively
short interval on the stable branch $S_{s}$, chosen as an initial
approximation. Figure~\ref{chaos}(b) shows how the connected
forward images of this interval become non-smooth, generating
growing wiggles around the unstable set ${\cal S}_{u}$. Note that
the unstable set ${\cal S}_{s}$ is easily computed using inverse
mapping , see Appendix~A. The insert in Fig.~\ref{chaos}(b) shows
the zoomed-in wiggles. One can see from Fig.~\ref{chaos} (a) and
(b) that upon entering the wiggling area, the phase point can land
back in the stable set ${\cal S}_{s}$, thereby completing another
round of subthreshold oscillations, or jump up to make a spike.
Such a situation is referred to as dynamical
uncertainty~\cite{book}. Loosely speaking, the place on ${\cal
S}_{u}$ where the wiggles become noticeable indicates the threshold
beyond which the behavior of the map becomes uncontrollable.

\begin{figure}[ht]
\centering\includegraphics[scale=0.7]{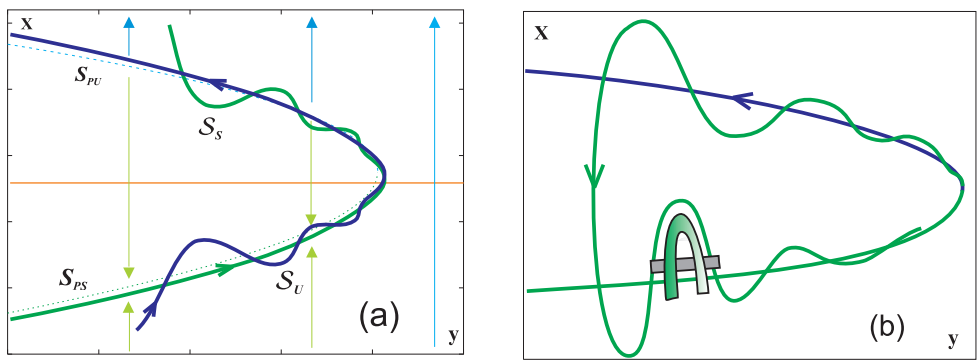}
\caption{Heteroclinic-like crossings of the critical sets ${\cal S}^s$
and ${\cal S}^u$. Self-crossings of the set ${\cal S}^s$, which
are one of the features of noninvertible maps, may generate a
topological Smale horseshoe.}\label{cross}
\end{figure}

This situation is similar to the case of a periodically driven
pendulum near a homoclinic orbit associated with the saddle point
at the top. Such a system is typically studied using a Poincar\'e
return mapping defined over the period of the external force. Under
proper conditions, the saddle fixed point of the mapping will
possess transversal crossings between its stable and unstable sets
(formerly, the stable and unstable separatrices of the saddle
equilibrium in the autonomous system). These crossings generate the
Smale horseshoe and, therefore, symbolic shift-dynamics in the
system, see details in \cite{arnold94,mira,synchro} and references
therein. By construction, this type of Poincare\'e mapping is a
diffeomorphism. However, our map~(\ref{map1}) is an endomorphism,
e.g. a non-invertible one, and as a result it may possess some
exotic features prohibited in the invertible maps, see more in
\cite{mira}. One of those features is that the set ${\cal S}_{s}$
may self-cross (so does the set ${\cal S}_{u}$ in the backward
time). This situation is sketched in Fig.~\ref{cross}(a) and also
can be seen on the trajectories' behavior in the lower left corner
of the attractor shown in Fig.~\ref{chaos}(a). One can assume that
these self-crossings can stimulate the conditions for the onset of
a topological Small horseshoe (see Fig.~\ref{cross}(b)), whose
presence  is a de-facto proof of complex chaotic dynamics.

\section{Conclusion}\label{Section7}

It is shown that a simple map can be employed to replicate the
behavior of neurons with self-sustained subthreshold oscillations.
These oscillations are achieved by a special selection of a
nonlinear function in the fast subsystem. The dynamical mechanisms
behind the transitions from silence to subthreshold oscillations
of small amplitude and then to spiking activity are explained
using the bifurcation approach.

Here we focused mostly on the individual dynamics of the map-based
model. As a result we considered only the case when $\beta$ and
$\sigma$ are constants. One may notice from (\refmap) that the
parameter $\beta$ can be eliminated by using the variable
transformation $y+\beta \mapsto y$. However, the role of input
parameter $\beta$ becomes important when a time dependant input is
considered, see~\cite{Rulkov2002} for detail. We would like to note
that for studies of non-autonomous dynamics of this map model one
needs to modify function (\ref{func}) to insure that no trajectory
of (\refmap) gets locked in the interval $0<x<y+1+
\beta$ as $\beta$ or $y$ increase. The discussion on suggested
alterations of the function to resolve this problem can be found
elsewhere~\cite{Rulkov2002}.

\section{Acknowledgment}\label{Section8}

N.R. was supported in part by U.S. Department of Energy (grant
DE-FG03-95ER14516). A.S. acknowledges the RFBR grants
No.~02-01-00273 and No.~01-01-00975.

\section{Appendix: Inverse map}\label{app1}

To locate the unstable set ${\cal S}^u$ which is a  surface of
slow motion  one should consider the inverse map $F_{inv} :
(x,y)\mapsto (\bar x, \bar y) \in D$ defined in $D:=\, \{
-1-\alpha/2 \le x \le 0\}$. Within $D$ the inverse $F_{inv}$
assumes the following form
 \setcounter{eq}{\value{equation}} \addtocounter{eq}{+1}
\setcounter{equation}{0}
\renewcommand{\theequation}{\theeq \alph{equation}}
\begin{eqnarray}
\bar x &=& \alpha x +(x+1)^2 +y+\beta, \label{mapxD}\\
\bar y &=& y- \mu (x+1-\sigma). \label{mapyD}
\end{eqnarray}
\newcommand{\Dmap}{1}
\setcounter{equation}{\value{eq}}
\renewcommand{\theequation}{\arabic{equation}}
Subtracting (\ref{mapyD}) from (\ref{mapxD}) one gets
\begin{eqnarray}\label{subsr1}
\bar x - \bar y = \alpha x +(x+1)^2 +\mu (x+1-\sigma)+\beta.
\end{eqnarray}
Solving it for $x$ one can find $x$ as a the following function in
$\bar x$ and $\bar y$
\begin{eqnarray}\label{invx}
x=-(2+\alpha+\mu)/2 + \sqrt{ (2+\alpha+\mu)^2/4 -(1+\beta+ \mu
(1-\sigma)-\bar x +\bar y}.
\end{eqnarray}
To get the equation for the $y$-variable expression (\ref{invx})
must be plugged  into (\ref{mapyD}).

\end{document}